\documentclass[aps,twocolumn,superscriptaddress,numerical,showpacs]{revtex4-1}

\usepackage{slashed}
\usepackage{epsfig}
\usepackage{amsmath}
\usepackage{amssymb}
\usepackage{graphicx}
\usepackage{graphicx}
\usepackage{dcolumn}
\usepackage{bm}
\usepackage{bbold}
\usepackage{hyperref}
\hypersetup{
     colorlinks   = true,
     citecolor    = blue,
     urlcolor = blue,
     linkcolor = blue
}

\begin{document}

\title{Elastic Gauge Fields and Hall Viscosity of Dirac Magnons}

\author{Yago Ferreiros}
  \affiliation{Department of Physics, KTH Royal Institute of Technology, SE-106 91 Stockholm, Sweden}
  \email{ferreiros@kth.se}
\author{Mar\'ia A. H. Vozmediano}
  \affiliation{Instituto de Ciencia de Materiales de Madrid,\\
CSIC, Cantoblanco; 28049 Madrid, Spain.}

\date{\today}

\begin{abstract}
We analyze the coupling of elastic lattice deformations to the magnon degrees of freedom
of magnon Dirac materials. For a Honeycomb ferromagnet we find that, as it happens in the case of graphene, elastic gauge fields appear coupled to the magnon pseudospinors. For deformations that induce constant pseudomagnetic fields, the spectrum around the Dirac nodes splits into pseudo-Landau levels.
We show that when a Dzyaloshinskii-Moriya  interaction is considered, a topological gap opens in the system and a Chern-Simons effective action for the elastic degrees of freedom is generated. Such a term encodes a phonon Hall viscosity response, entirely generated by quantum fluctuations of magnons living in the vicinity of the Dirac points. The magnon Hall viscosity vanishes at zero temperature, and grows as temperature is raised and the states around the Dirac points are increasingly populated.
\end{abstract}

\pacs{Valid PACS appear here}
\maketitle


\section{Introduction}

Since the synthesis of graphene, Dirac materials are a common trend in condensed matter \cite{WBB14}. They appear in a variety of compounds with the  characteristic that their low energy elementary excitations are described by a Dirac Hamiltonian. Most of these materials 
are based on electrons moving on a periodic lattice and their elementary excitations are  fermionic degrees of freedom. Interesting exceptions are the optical lattices \cite{JMetal14}, photonic crystals \cite{HLetal11}, or lattice constructions like the artificial graphene described in  \cite{KWM12}. Magnons are becoming interesting players in the new condensed matter systems.  The proposal of topological magnons \cite{ZLetal13,MHM14,PYetal15,CHFS15,NKKL17,LK17}  has been followed by that of magnon Dirac matter  \cite{FBB15,LLetal16,Owerre16,KOetal16,PBB17,SW17,O17,LLH17,JN17,O417}. Dirac magnons have been very recently observed in three-dimensional Dirac antiferromagnets \cite{BWW17}. Although described by a Dirac Hamiltonian, the constituent elements are bosons with no electric charge. In two dimensions, Dirac magnons can arise as excitations of Honeycomb lattices of localized spins. Honeycomb ferromagnets are realized in chromium trihalides $\mathrm{CrX_3}$ (X = F, Cl, Br, I) \cite{PBB17}, with recent theoretical and experimental works demonstrating the robustness of the Honeycomb ferromagnetic phase in $\mathrm{CrI_3}$ \cite{GLLZ17,HCJX17,LR17}. In addition, they can be artificially engineered by depositing magnetic impurities on metallic substrates \cite{KOetal16}.

An interesting question is what properties of "Dirac physics" will survive in these bosonic constructions and what will be the physical interpretation of the "spinor" response functions. Irrespective of the nature of the elementary excitations, the solutions of the Dirac equation will be spinors in a general sense. In the particular case of the Honeycomb ferromagnet lattice constructions, the pseudospin is associated to the sublattice degree of freedom. Still, the wave function will have a non trivial Berry phase in momentum space and may acquire topological properties when gapped, leading to magnon responses such as the thermal Hall effect and the spin Nernst effect \cite{KNL10,OIKNT10,HCLO15,O216,KOetal16,COX16,ZK16,MHM16,O317,STS17,NKL17,ZK17,RBD17}.

Historically, the most interesting properties of the Dirac Hamiltonian in two spatial dimensions have been associated with the responses to vector fields minimally coupled to the spinors, in particular the electromagnetic field. A very appealing phenomenon in graphene and similar Dirac materials   is  the  coupling of the lattice deformations to the spinorial degrees of freedom. This led to the prediction of elastic gauge fields in two and three-dimensional Dirac materials, such as graphene \cite{VKG10} and Weyl semimetals \cite{CFLV15}. In the case of graphene, the presence of elastic gauge fields was confirmed experimentally in \cite{LBetal10} -see also the realization in \cite{KWM12}). The general influence of elastic lattice deformations on the spinor degrees of freedom was analyzed in \cite{MJSV13} and has been reviewed recently in \cite{Amorim16}.  In this work we study the influence of lattice deformations on the magnon physics   in the magnon Dirac materials. The main aspect of this work is related to the vector coupling of elasticity  with the magnons. As an immediate consequence, under the particular strain giving rise to constant pseudomagnetic fields, the magnon spectrum will be organized in ``magnon Landau levels", an unexpected issue that can be probed experimentally. 

\begin{figure*}
(a)
\begin{minipage}{.4\linewidth}
\includegraphics[scale=0.53]{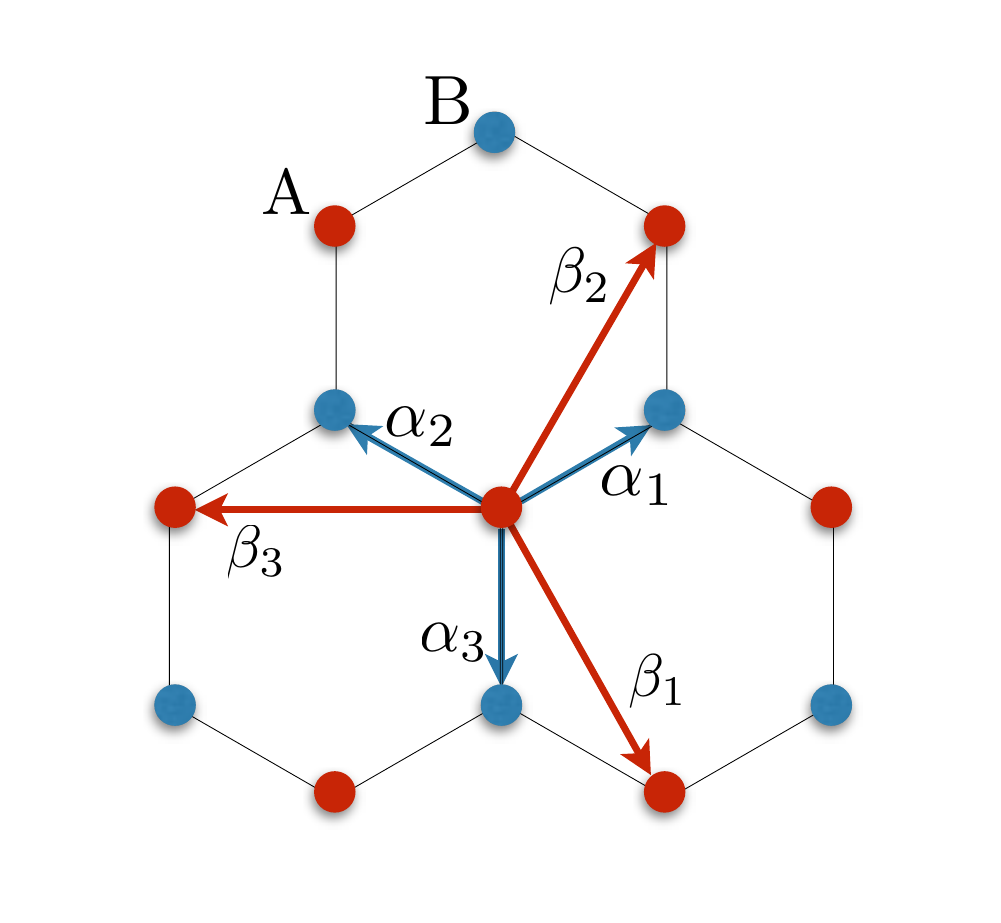}
\end{minipage}
(b)
\begin{minipage}{.4\linewidth}
\includegraphics[scale=0.5]{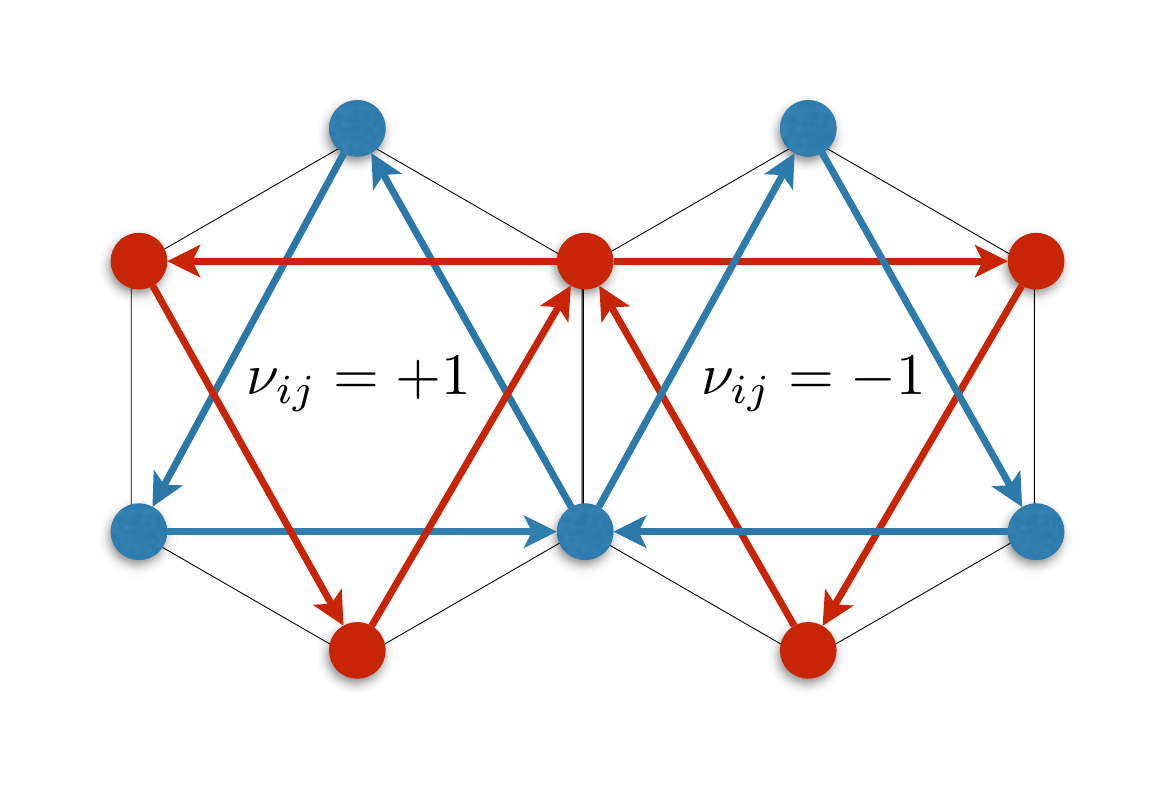}
\end{minipage}

\caption{(a) Honeycomb lattice with the sites corresponding to the two sublattices A and B represented by red and blue circles, respectively. The nearest and next-nearest neighbor vectors, $\bm\alpha_i$ and $\bm\beta_i$, are represented by blue and red arrows, respectively. (b) Relative sign $\nu_{ij}$ of the Dzyaloshinskii-Moriya interaction.}
\label{fig lattice}
\end{figure*}

Furthermore, when a Dzyaloshinskii-Moriya interaction \cite{D58,M60,KNB05} for spins is considered, a topological gap opens and the Dirac magnon system resembles a Haldane model \cite{H88} for magnons \cite{Owerre16,KOetal16}. For fermionic excitations, the Haldane model has been recently shown to generate a phonon Hall viscosity response when coupled to elasticity, arising from a Chern-Simons term for the elastic gauge fields \cite{CFLV16}. The Hall viscosity is a topological viscoelastic response first described in the context of quantum Hall fluids\cite{A95}, and subsequently investigated in a number of works \cite{R09,H09,RR11,HS12,HLF11,H14,SHR15} (please note that this list is just a small sample and not exhaustive). In this work we show that for magnon excitations, the Haldane model presents a Hall viscosity response which increases with temperature and vanishes at $T=0$. We call it magnon Hall viscosity as it is entirely generated by magnon fluctuations.

\section{The model}
We consider a ferromagnetic material whose localized spins are arranged on a Honeycomb lattice. The corresponding model Hamiltonian reads
\begin{equation}
H=-J\sum_{\langle i,j\rangle}\bm{S}_i\cdot\bm{S}_j-B\sum_iS_i^z,
\label{eq model no DM}
\end{equation}
where the first term represents the isotropic Heisenberg interaction ($J > 0$) between nearest neighbors, and the second term is a Zeeman coupling to a magnetic field applied along the $z$ direction. By applying the Holstein-Primakoff transformation
\begin{gather}
S_i^+=S_i^x+iS_i^y=\sqrt{2S-n_i}\,\,d_i,\\
S_i^-=(S_i^+)^\dagger,\quad S_i^z=S-n_i,
\end{gather}
with $n_i=d_i^\dagger d_i$, we arrive at the following effective magnon Hamiltonian
\begin{equation}
H=(3JS+B)\sum_i d_i^\dagger d_i-JS\sum_{\langle i,j \rangle}(d_i^\dagger d_j+\mathrm{H.c.}).
\label{eq trivial magnon Hamiltonian}
\end{equation}
up to second order in the magnon operators. Although higher order corrections may account for anharmonic magnon effects, they preserve the relevant symmetries and therefore do not destroy the Dirac magnon properties \cite{FBB15}. Fourier transforming the Hamiltonian we get \cite{FBB15,Owerre16,KOetal16}
\begin{equation}
H=\sum_{\bm k\in B.Z.}\Psi_{\bm k}^\dagger[(3JS+B)\,\mathbb{1}+\bm d(\bm k)\cdot\bm\sigma]\Psi_{\bm k},
\label{eq Ham Fourier}
\end{equation}
where $\Psi_{\bm k}=(a_{\bm k},b_{\bm k})$, $a$ and $b$ are the magnon annihilation operators on the two sublattices, $\bm\sigma=(\sigma_x,\sigma_y,\sigma_z)$, and
\begin{equation}
\bm d(\bm k)=\sum_i
\begin{pmatrix}
-JS\cos[\bm k\cdot\bm{\alpha}_i]\\
JS\sin[\bm k\cdot\bm{\alpha}_i]\\
0
\end{pmatrix},
\end{equation}
with the vectors $\bm \alpha_i$ defined in Fig. \ref{fig lattice}(a). The dispersion relations for the upper and lower energy bands are given by
\begin{equation}
E^\pm(\bm k)=3JS+B\pm|\bm d(\bm k)|.
\end{equation}
The upper and lower bands meet at six Dirac points, with only two of them being inequivalent. Let us pick $\bm K_s=(s4\pi/3\sqrt{3}a,0)$, with $s=\pm$ and where $a$ is the distance between nearest neighbors. Expanding around these two Dirac points we get
\begin{equation}
\bm d_D^s(\bm k)=\begin{pmatrix}
s v_Jk_x/2\\
-v_Jk_y/2\\
0
\end{pmatrix},
\end{equation}
with $v_J=3JSa/2\hbar$. The continuum Dirac Hamiltonian is then
\begin{equation}
H_D^s=\int \frac{d^2 k}{(2\pi)^2}\,\Psi_{\bm k}^\dagger\,\hbar \,v_J(sk_x\sigma_x-k_y\sigma_y)\Psi_{\bm k}.
\end{equation}

\begin{figure*}
(a)
\begin{minipage}{.4\linewidth}
\includegraphics[scale=0.5]{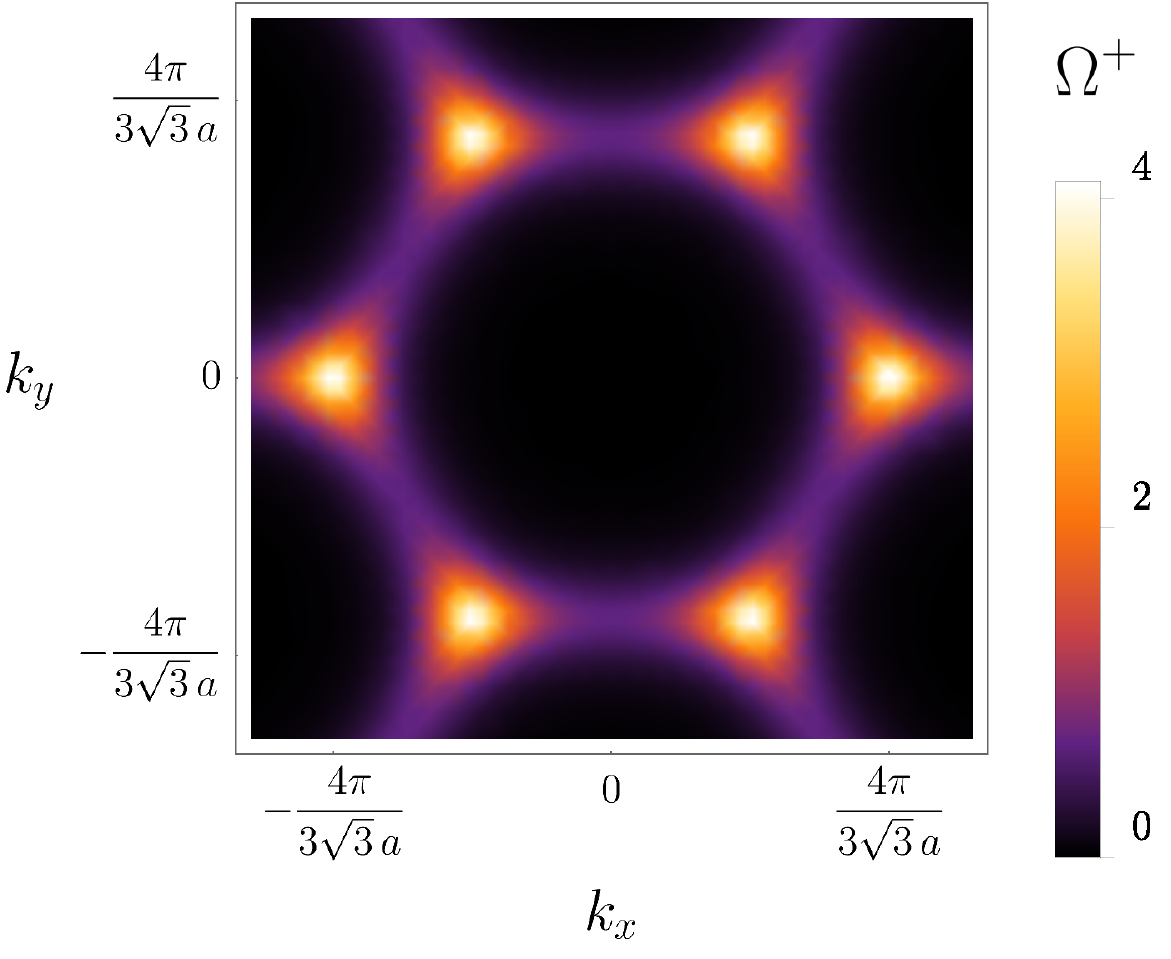}
\end{minipage}
(b)
\begin{minipage}{.4\linewidth}
\includegraphics[scale=0.5]{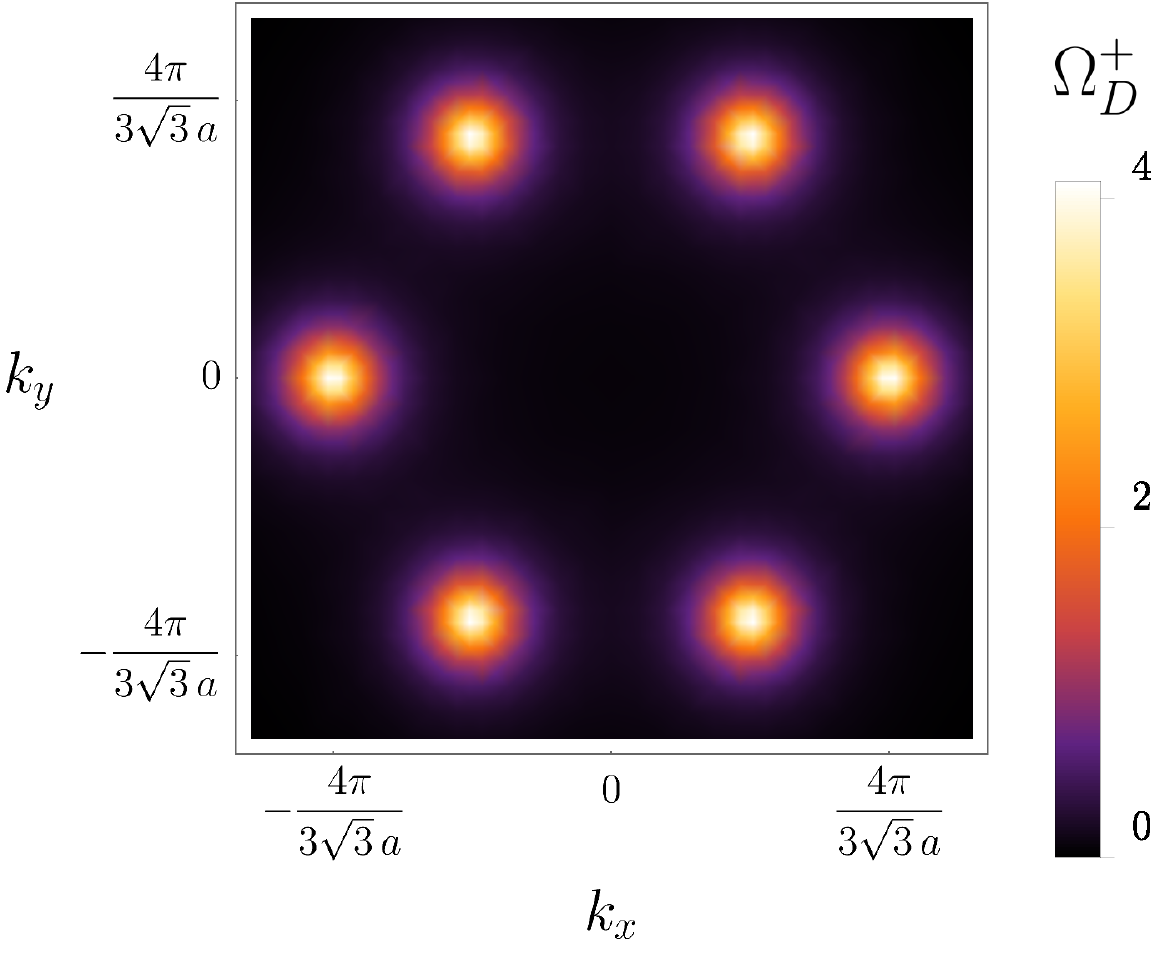}
\end{minipage}

\caption{Total Berry curvature (a) as computed from the vector $\bm d_{DM}$, versus Dirac Berry curvature (b), obtained by adding the Berry curvatures of the six gapped Dirac cones located at the six Dirac points.}
\label{fig Berry}
\end{figure*}

\section{Elastic gauge fields and magnon Landau levels}
To introduce elastic deformations of the lattice, we allow for small displacements of the spins away of their initial positions. We can write $J\equiv J_i$, where $i$ stands for the three nearest-neighbors. Now the value of the exchange coupling changes according to variations of the positions of the spins. Applying the same methodology as for graphene \cite{VKG10,K12}, expanding the exchange coupling around its equilibrium value $J$
\begin{equation}
J_i=J-\frac{\beta J}{a^2}\,\bm\alpha_i\cdot\delta\bm{u}_i,
\label{eq expansion J}
\end{equation}
where $\delta\bm{u}_i$ is the displacement of the spins around their equilibrium configurations, and $\beta$ is a parameter that controls the expansion and has to be computed from the microscopic properties of the system, we obtain the elastic vector fields
\begin{equation}
A_x^{el}=\frac{\beta JS}{v_J}(u_{xx}-u_{yy}),\label{eq A1}
\end{equation}
\begin{equation}
A_y^{el}=-\frac{2\beta JS}{v_J}u_{xy}.\label{eq A2}
\end{equation}
Here $u_{ab}=(\partial_au_b+\partial_bu_a)/2$ is the strain tensor, $a,b = x,y$, and $\bm{u}$ is the displacement vector, which arises when taking the continuum limit $\delta\bm{u}_i\sim(\bm\alpha_i\cdot\bm\partial)\,\bm u$ \cite{VKG10,K12}. The continuum Dirac Hamiltonian now reads
\begin{gather}
H_D^s=\int \frac{d^2 k\,d^2k'}{(2\pi)^4}\,\Psi_{\bm k}^\dagger\,v_J\Big\{[s\hbar k_x-A_x^{el}(\bm k-\bm k')]\sigma_x\nonumber\\
-[\hbar k_y-sA_y^{el}(\bm k-\bm k')]\sigma_y\Big\}\Psi_{\bm k'}.
\label{eq Ham Dirac elastic}
\end{gather}

One could engineer Dirac magnons using STM, by depositing magnetic defects on a metallic substrate forming a Honeycomb lattice of spins. For large separation between the spins, the exchange coupling between different sites can be written as $J_{i}\sim\sin(2k_F|\bm R_{i}|)/|\bm R_{i}|^2$ \cite{KOetal16}, as described by the RKKY interaction. Here $k_F$ is the Fermi wave vector of the substrate and $|\bm R_{i}|$ is the distance between nearest neighboring spins. Allowing for small displacement around the equilibrium configuration $\bm R_i=\bm\alpha_i+\delta\bm u_i$, we can expand $J_i$ to first order in $\delta\bm u_i$ and obtain Eq. (\ref{eq expansion J}) with $\beta$ given by
\begin{equation}
\beta=2-2ak_F\cot(2ak_F),
\label{eq:beta}
\end{equation}
and with $J\sim\sin(2ak_F)/a^2$.

By playing with the position of the spins on the substrate, one can engineer specific forms of $A_{x,y}^{el}$. A constant pseudomagnetic field, of opposite sign in the two valleys $s=\pm$, can be obtained by the following arrangement of the spins \cite{GKG10}
\begin{equation}
u_x=2qxy,\quad u_y=q(x^2-y^2),
\end{equation}
where $q$ is a parameter denoting the strength of the strain, with units of inverse length. This exact arrangement was experimentally realized for molecular graphene \cite{KWM12}. The value of the pseudomagnetic field associated to this distortion is
\begin{equation}
|eB|=\frac{8\beta J q}{v_J}=\frac{16\hbar\beta q}{3a}.
\label{eq pseudomagnetic field}
\end{equation} 
This is, magnons react to the distortion of the lattice in a similar way as particles of charge $e$ would react to a magnetic field of magnitude given by Eq. (\ref{eq pseudomagnetic field}). Solving the Dirac equation in the presence of this constant pseudomagnetic field gives rise to the quantized spectrum
\begin{equation}
E_n=\pm\sqrt{2\hbar |eB|\,v_J^2\,n},
\end{equation}
so that the magnons arrange into Landau levels around the Dirac points. Since the pseudomagnetic field is of opposite sign at the two valleys, there are counterpropagating edge states \cite{GKG10} and the total Chern number is zero. Finally we can estimate the size of the gap between Landau levels as a function of $J$. Taking a lattice spacing of $a\sim10\,\textrm{\r{A}}$ and a maximum strain strength of $q\sim10^{-3}\,\textrm{\r{A}}^{-1}$ \cite{KWM12}, setting $S=1$, and doing $ak_F\ll1$ in Eq. (\ref{eq:beta}) so that $\beta\sim1$, we obtain $E_n\sim\pm\sqrt{n}J/2$. For the artificial Honeycomb lattice of magnetic defects we are considering, the value of $J$ will depend on the exchange interaction between the localized spins and the spin density of the Fermi sea, and on the distance between nearest neighboring spins \cite{KOetal16}.

\section{Magnon Hall viscosity}
Let us consider an extension of the model of Eq. (\ref{eq model no DM}) by adding an inversion symmetry breaking Dzyaloshinskii-Moriya (DM) term \cite{Owerre16,KOetal16}
\begin{gather}
H_{DM}=-J\sum_{\langle i,j\rangle}\bm{S}_i\cdot\bm{S}_j-B\sum_iS_i^z\nonumber\\
+D\sum_{\langle\langle i,j\rangle\rangle}\nu_{ij}\,\hat{\bm z}\cdot(\bm S_i\times\bm S_j).
\label{eq model DM}
\end{gather}
The DM interaction (last term) depends on the relative position of two next-nearest neighboring spins through the constants $\nu_{ij}=-\nu_{ij}=\pm1$ [Fig. \ref{fig lattice}(b)]. The effective Hamiltonian after the application of the Holstein-Primakoff transformation turns out to be
\begin{gather}
H_{DM}=(3JS+B)\sum_i d_i^\dagger d_i-JS\sum_{\langle i,j \rangle}(d_i^\dagger d_j+\mathrm{H.c.})\nonumber\\
-DS\sum_{\langle\langle i,j\rangle\rangle}(i\nu_{ij}d_i^\dagger d_j+H.c.),\label{eq magnon ham DM}
\end{gather}
which basically is the Haldane model for magnons proposed in Refs. \cite{Owerre16,KOetal16}. 

The Fourier transform of Hamiltonian (\ref{eq magnon ham DM}) is captured by Eq. (\ref{eq Ham Fourier}) with the vector $\bm d$ given now by
\begin{equation}
\bm d_{DM}(\bm k)=\sum_i
\begin{pmatrix}
-JS\cos[\bm k\cdot\bm{\alpha}_i]\\
JS\sin[\bm k\cdot\bm{\alpha}_i]\\
2DS\sin[\bm k\cdot\bm\beta_i]
\end{pmatrix},
\end{equation}
with the next-nearest vectors $\bm \beta_i$ defined in Fig. \ref{fig lattice}(a). The dispersion of the upper and lower energy bands is
\begin{equation}
E_{DM}^\pm(\bm k)=3JS+B\pm|\bm d_{DM}(\bm k)|,
\end{equation}
with energy gap at the Dirac points, $\bm K_s$, of $|\Delta|=6\sqrt{3}DS$. In analogy with the trivial honeycomb magnon system of Eq. (\ref{eq trivial magnon Hamiltonian}), when coupled to elasticity and expanded around the Dirac points, the magnon Haldane model gives rise to the elastic gauge fields of Eqs. (\ref{eq A1}) and (\ref{eq A2}). The topological nature of the Hamiltonian of Eq. (\ref{eq magnon ham DM}) is captured by the Berry curvatures of the upper and lower magnon bands \cite{XCN10}
\begin{gather}
\Omega^\pm(\mathbf{k})=\mp\frac{1}{2}\hat{\bm d}\cdot(\partial_{k_x}\hat{\bm d}\times\partial_{k_y}\hat{\bm d}),\\
\hat{\bm d}=\frac{\bm d_{DM}}{|\bm d_{DM}|},
\end{gather}
with Chern numbers
\begin{equation}
C^\pm=\frac{1}{2\pi}\int_{\mathrm{B.Z.}}\Omega^\pm\,d^2k=\pm1.
\end{equation}
One may wonder what is the physical picture when pseudomagnetic fields coexist with a DM topological gap. If we look at the Landau level spectrum, we get \cite{}: $E_n=\pm(2\hbar |eB|\,v_J^2\,n+\Delta^2)^{1/2}$ for $n\geq1$, and $E_0=\mathrm{sign}(eB)\Delta$ for the lowest Landau level. In one valley the lowest Landau level is lowered in energy, while in the other valley it is raised. Regarding the spectrum at the boundary, due to the non-zero Chern number there will be a chiral edge state coexisting with the counterpropagating edge states due to the psudomagnetic field. The spectrum in the presence of pseudomagnetic fields and non-zero Chern number has been discussed in detail in two \cite{HH13} and three- \cite{GVVI16,PCF16} dimensional systems (spin-orbit coupled graphene and time-reversal breaking Weyl semimetals, respectively).

In what follows, the coupling to elasticity will be treated perturbatively, with the equilibrium configuration given by the undeformed Honeycomb lattice. Under minimal coupling to gauge fields, all quantum fluctuations in bands with non zero Chern number contribute to the generation of Cherns-Simons terms. As an example, electronic systems with non zero Chern number, this is fermionic Chern insulators, give rise to the quantum anomalous Hall effect, which at the level of the action is given by a Chern-Simons term for the electromagnetic field. In our case, the minimal coupling is to elasticity, so quantum fluctuations in our Haldane model for magnons (which basically is a bosonic Chern insulator) will give rise to a Chern-Simons term for the elastic gauge fields. The mechanism is analogous to what happens in the Haldane model of electrons coupled to elasticity \cite{CFLV16}, and the Chern-Simons term reads
\begin{equation}
S_{CS}=\frac{\tilde{C}}{4\pi\hbar}\int d^3x\,\epsilon^{\mu\rho\nu}A_\mu^{el}\partial_\rho A_\nu^{el},
\label{eq CS}
\end{equation}
where $\mu,\rho,\nu=t,x,y$, $A^{el}_t=0$, and $\tilde{C}$ is given by the contributions to the Chern number of the occupied bands. The difference with the electronic case, which at half filling is given by $\tilde{C}=C^-=-1$, is just that magnons are bosons, so they contribute with the Bose-Einstein distribution function
\begin{equation}
\tilde{C}=\frac{1}{2\pi}\sum_{\tau=\pm}\int_{\mathrm{B.Z.}}\Omega^{\tau}(\mathbf{k})\rho^\tau(\mathbf{k})\,d^2k,
\label{eq C minimal coupling}
\end{equation}
with
\begin{equation}
\rho^\tau(\mathbf{k})=\frac{1}{e^{E_{DM}^\tau(\mathbf{k})/\kappa T}-1}.
\end{equation}
However, Eq. (\ref{eq C minimal coupling}) is only valid if $A_\mu^{el}$ couples minimally to magnons in the entire Brillouin zone. This is not the case though, as the coupling to elasticity is only minimal near the Dirac points, where the physics of the system is dominated by the Dirac approximation, so strictly speaking Eq. (\ref{eq C minimal coupling}) is not totally accurate. By moving away from the Dirac points, the ``Diracness" is lost and magnons do not contribute to Eq. (\ref{eq CS}) anymore. To take this into account we shall define a Dirac Berry curvature $\Omega_D$, i. e. a Berry curvature that only captures the Dirac contributions. It can be constructed by linearizing $\bm d_{DM}$ around all six Dirac points, and adding the Berry curvatures of the six resulting gapped Dirac cones. As a check, it is apparent by comparison of $\Omega$ and $\Omega_D$ [Figs. \ref{fig Berry}(a) and \ref{fig Berry}(b)], that such a Dirac Berry curvature is missing the information coming from the ``non-Diracness" of the bands.

\begin{figure}
\includegraphics[scale=0.63]{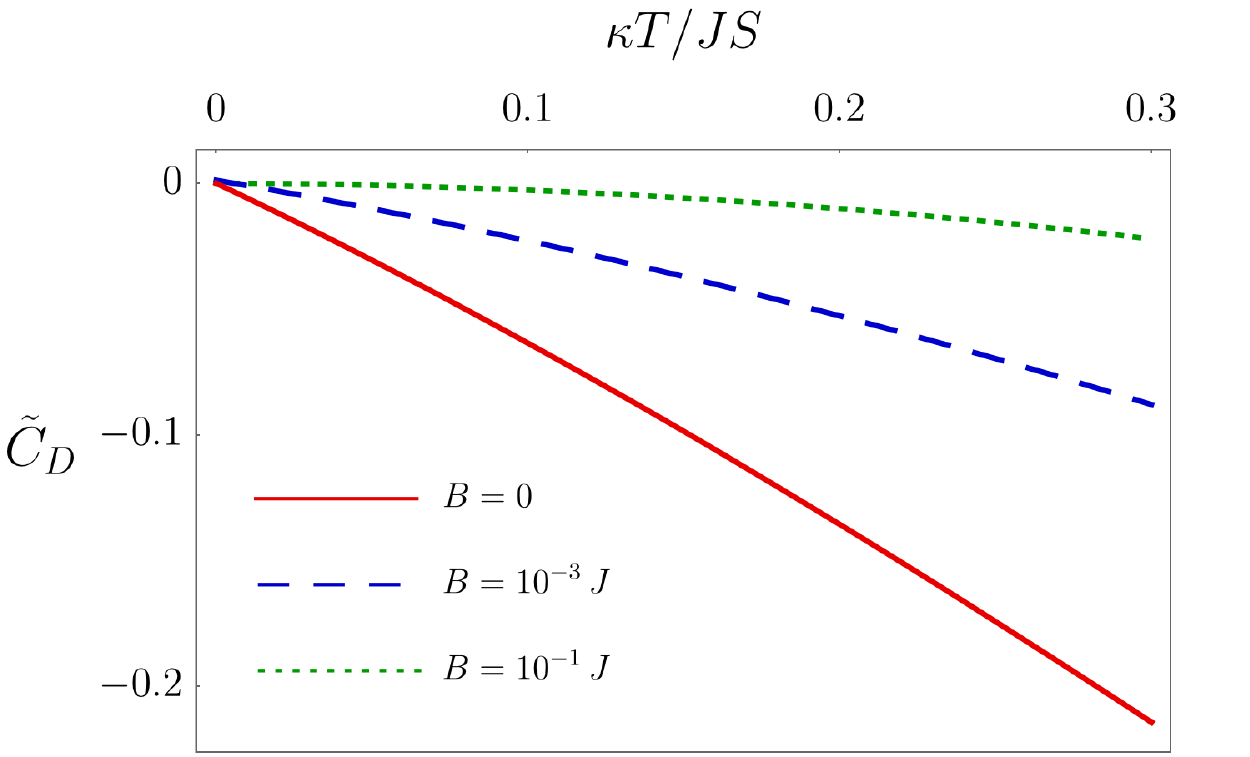}
\caption{$\tilde{C}_D$ as a function of temperature for a value of the DM interaction of $D=J/10$ and for three different values of the magnetic field.}
\label{fig Chern magnons}
\end{figure}

The correct value of $\tilde{C}$ can therefore be obtained by substituting $\Omega$ by $\Omega_D$ in Eq. (\ref{eq C minimal coupling})
\begin{equation}
\tilde{C}_D=\frac{1}{2\pi}\sum_{\tau=\pm}\int_{\mathrm{B.Z.}}\Omega_D^{\tau}(\mathbf{k})\rho^\tau(\mathbf{k})\,d^2k,
\label{eq C}
\end{equation}
so that the Chern-Simons term becomes
\begin{equation}
S_{CS}=\frac{\tilde{C}_D}{4\pi\hbar}\int d^3x\,\epsilon^{\mu\rho\nu}A_\mu^{el}\partial_\rho A_\nu^{el}.
\label{eq CS Dirac}
\end{equation}
Inserting the explicit form of the elastic gauge fields, given by Eqs. (\ref{eq A1}) and (\ref{eq A2}), in the Chern-Simons term, we get
\begin{equation}
S_{CS}=\frac{4\hbar\beta^2\tilde{C}_D}{9\pi a^2}\int d^3x\,(u_{xx}-u_{yy})\dot{u}_{xy}.
\label{eq CS term final}
\end{equation}
The stress tensor can be obtained by differentiating the action with respect to the strain tensor
\begin{equation}
T_{ab}=-\frac{\delta S}{\delta u_{ab}},
\label{eq stress tensor definition}
\end{equation}
with $a,b=x,y$, while the viscosity tensor, $\eta$, characterizes the dependence of the stress tensor on the strain rate
\begin{equation}
T_{ab}=-\eta_{abcd}\,\dot{u}_{cd}.
\end{equation}
$\eta$ is symmetric under the interchange of $a$ and $b$, and $c$ and $d$, due to the symmetry of the strain tensor. The Hall viscosity, $\eta^H$, is given by the antisymmetric part of $\eta$ under the interchange of the pairs $a,b$ and $c,d$ \cite{A95}. For an isotropic 2D system, $\eta^H$ has only one independent component: $\eta_{xxxy}=-\eta_{xyxx}=-\eta_{yyxy}=\eta_{xyyy}=\eta^H$ \cite{A95}. Then, by applying Eq. (\ref{eq stress tensor definition}) to the Chern-Simons action (\ref{eq CS term final}) we get
\begin{equation}
\eta^H=-\frac{4\hbar\beta^2\tilde{C}_D}{9\pi a^2}.
\end{equation}

In Fig. \ref{fig Chern magnons} we plot $\tilde{C}_D$ as a function of $\kappa T/JS$, for a DM interaction of $D=J/10$. At zero temperature, a Bose-Einstein condensate forms at zero energy, where the Berry curvature vanishes and consequently $\tilde{C}_D=0$. As the temperature increases, states with non zero Berry curvature are populated, and $\tilde{C}_D$, and by extension $\eta^H$, become finite. The magnitude of $\eta^H$ continues to grow as the population around the Dirac points, where the Berry curvature is maximal, increases, until the temperature rises to a point where the magnon picture is no longer valid. The magnon description is reasonable for temperatures considerably smaller than the Curie temperature $T_c$, above which the system is disordered by thermal fluctuations. If we take a monolayer of $\mathrm{CrI_3}$ as a candidate material hosting Dirac magnons, its Curie temperature is $T_c=45\,\textrm{K}$ \cite{HCJX17} and its exchange coupling and spin have been estimated to be $J=2.2\,\textrm{meV}$ and $S=3/2$ \cite{LR17}. If we assume the validity of the magnon picture up to a temperature of around $T\approx T_c/10=4.5\,\textrm{K}$, our calculations are valid up to $\kappa T/JS\approx0.12$, which corresponds to a maximum value of roughly $\tilde{C}_D\sim0.1$. For higher temperatures, an alternative picture based on the Schwinger-boson representation of spin can be invoked to correctly capture the topological properties of the system \cite{KOetal16}. As a final note, we see that the application of a magnetic field increases the energies of magnons, which leads to a decrease of the Hall viscosity.

\section{Conclusions and discussion}
We have studied the influence of lattice deformations on the magnon physics of a Honeycomb ferromagnet. We have proven that, in the vicinity of the Dirac points, elasticity couples at lowest order as vector fields to Dirac magnons. For strain configurations giving rise to constant pseudomagnetic fields, magnons arrange in Landau levels. Such strain configurations can be realized  using STM, by depositing magnetic defects on a metallic substrate. The presence of Landau levels could be tested experimentally by using, among other techniques, inelastic neutron scattering \cite{BWW17}.

By including a Dzyaloshinskii-Moriya interaction, a topological gap opens and a Chern-Simons effective action for the elastic gauge fields is generated. Such a term encodes a phonon Hall viscosity response, which is generated entirely by quantum fluctuations of magnons living in the vicinity of the Dirac points. Its value vanishes at zero temperature, and grows as temperature is raised and the states around the Dirac points are increasingly populated. Having in mind that measuring the Hall viscosity is not an easy task, a possible natural direction is to measure changes in the phonon structure generated by $\eta^H$ \cite{SHR15,BCQ12,CFLV16}. 

We can compare the Hall viscosity of the Haldane model for magnons obtained here to that of the Haldane model for electrons (with chemical potential inside the gap) computed in \cite{CFLV16}, arising also from elastic gauge fields. In the electronic case, the Hall viscosity is $\eta^H_{elec}=4\hbar\beta^2/9\pi a^2$, whereas for magnons we obtained $\eta^H_{mag}=-4\hbar\beta^2\tilde{C}_D/9\pi a^2$. Even if they are material dependent, we can reasonably assume $\beta$ and $a$ to be roughly of the same order in magnon and electronic systems, so we have
\begin{equation}
\frac{\eta^H_{mag}}{\eta^H_{elec}}\sim-\tilde{C}_D.
\end{equation}
If we take our Dirac magnon system to be a monolayer of $\mathrm{CrI_3}$, as we estimated above our calculations are valid up to temperatures of $4.5\,\textrm{K}$, for which we obtain a maximum value of $\tilde{C}_D\sim0.1$. Then we get
\begin{equation}
\frac{\eta^H_{mag}}{\eta^H_{elec}}\sim0.1,
\end{equation}
which means that, at best, the magnon Hall viscosity is one order of magnitude down on its electronic counterpart.

It is important to remark that, in electronic systems, the Hall viscosity arising from elastic gauge fields is several orders of magnitude bigger than that coming from conventional phonons or metric deformations \cite{CFLV16}, which greatly increases the chances of experimental detection. Basically, the conventional Hall viscosity under magnetic fields is inversely proportional to the squared magnetic length, $l_B$, whereas in the Hall viscosity coming from elastic gauge fields the magnetic length is substituted by the lattice spacing $a$ \cite{CFLV16}. For magnetic fields of the order of $10$ Tesla, and for the lattice constant of, say, graphene ($a\approx2.5\,\textrm{\r{A}}$), we have $l_B^2/a^2\sim 10^{3}$, so the value of the Hall viscosity from elastic gauge fields is three orders of magnitude bigger than the conventional one. Therefore, even if the magnon Hall viscosity computed here is (at best) one order of magnitude down on its electronic counterpart, it is still (at best) two orders of magnitude bigger than the conventional Hall viscosity of electrons under magnetic fields.

\acknowledgments
We thank Hector Ochoa and Alberto Cortijo for enlightening and fruitful discussions. Y. F. acknowledges support from the ERC Starting Grant No. 679722. The work of M. V.  has been supported by Spanish MECD grant FIS2014-57432-P, the Comunidad de Madrid
MAD2D-CM Program (S2013/MIT-3007), and by the PIC2016FR6. 


%

\end{document}